\documentclass[lettersize,journal]{IEEEtran}
\usepackage{amsmath,amsfonts}
\usepackage{enumerate}
\usepackage{array}
\usepackage[caption=false,font=normalsize,labelfont=sf,textfont=sf]{subfig}
\usepackage{textcomp}
\usepackage{stfloats}
\usepackage{url}
\usepackage{verbatim}
\usepackage{graphicx}
\usepackage{cite}
\usepackage{balance}
\usepackage{amssymb}
\usepackage{algorithm,algpseudocode}

\allowdisplaybreaks[4]

\title{Semantic Transmission Framework in \\Direct Satellite Communications}

\author{Chong Huang, \IEEEmembership{Member, IEEE}, Xuyang Chen, Jingjing Cui, \IEEEmembership{Senior Member, IEEE}, Jingfu Li, \IEEEmembership{Member, IEEE},\\Pei Xiao, \IEEEmembership{Senior Member, IEEE}, Gaojie Chen, \IEEEmembership{Senior Member, IEEE}, Rahim Tafazolli, \IEEEmembership{Fellow, IEEE}
\thanks{The work presented in this article was supported by the U.K. Engineering and Physical Sciences Research Council under Grant EP/X013162/1. }
\thanks{C. Huang, P. Xiao and R. Tafazolli are with 5GIC \& 6GIC, Institute for Communication Systems (ICS), University of Surrey, Guildford, GU2 7XH, United Kingdom (e-mail: chong.huang@surrey.ac.uk; p.xiao@surrey.ac.uk; r.tafazolli@surrey.ac.uk).}
\thanks{J. Li and J. Cui are with School of Information Science and Technology, Southwest Jiaotong University, Chengdu, Sichuan, China (e-mail: jingfuli@swjtu.edu.cn, jingjing.cui@swjtu.edu.cn).}
\thanks{X. Chen is with the College of Electronics and Information Engineering, Shenzhen University, Shenzhen 518060, China. (e-mail: chenxuyang2021@email.szu.edu.cn)}
\thanks{G. Chen is with the School of Flexible Electronics (SoFE) \& State Key Laboratory of Optoelectronic Materials and Technologies, Sun Yat-sen University, Guangdong, China (e-mail: gaojie.chen@ieee.org).}
}

\begin{document}
\captionsetup[figure]{name={Fig.},labelsep=period}

\maketitle

\begin{abstract}
Insufficient link budget has become a bottleneck problem for direct access in current satellite communications. In this paper, we develop a semantic transmission framework for direct satellite communications as an effective and viable solution to tackle this problem. To measure the tradeoffs between communication, computation, and generation quality, we introduce a semantic efficiency metric with optimized weights. The optimization aims to maximize the average semantic efficiency metric by jointly optimizing transmission mode selection, satellite-user association, ISL task migration, denoising steps, and adaptive weights, which is a complex nonlinear integer programming problem. To maximize the average semantic efficiency metric, we propose a decision-assisted REINFORCE++ algorithm that utilizes feasibility-aware action space and a critic-free stabilized policy update. Numerical results show that the proposed algorithm achieves higher semantic efficiency than baselines.
\end{abstract}

\section{Introduction}
\IEEEPARstart{W}{ith} the rapid development of sixth-generation (6G), satellite communications have become an important avenue for future wireless networks due to their capability to provide seamless connectivity to global ground users \cite{9755278}. Compared with terrestrial communication systems, satellite communications can effectively overcome geographical limitations and guarantee service availability without ground infrastructure. Thus, it is particularly suitable for emergency communication scenarios. In recent years, the rapid deployment of satellite constellations such as Starlink and OneWeb, have significantly accelerated the construction of large-scale satellite communication networks and bring new frontiers for applications in cognitive communications \cite{10287142}, rate splitting multiple access (RSMA) \cite{10915662}, and edge computing \cite{10440193}.

However, due to limited transmit power, constrained bandwidth, and extremely long transmission distances, direct satellite communications face stringent link budget limitations \cite{10286242}. Long distance transmission between satellites and ground communication devices suffers from severe signal attenuation and propagation delay, spectrum efficiency becomes increasingly important for future satellite networks. To address this challenge, semantic communications, which aim to enable meaning-oriented information transmission, have emerged for future wireless communications in recent years \cite{10554663}. By understanding the semantics of the data, semantic communications extract semantic features and transmit them to the receiver, then the receiver can reconstruct the data based on these features. As a result, semantic communications can significantly reduce the required radio resources to enhance the spectral efficiency, rendering it particularly suitable for satellite communications under limited spectrum resource conditions \cite{11006980}.

Recently, many studies have investigated semantic communications in wireless networks \cite{10639525}. In \cite{10901733}, a deep learning-based semantic communication framework was proposed for image transmissions. To study the potential of semantic communications in satellite networks, a federated learning method was proposed to enhance the privacy of users in semantic communications for satellite edge computing in \cite{10445211}. In \cite{10960418}, an encoder and decoder design of semantic communications was studied to avoid retransmissions in satellite networks. However, the tradeoffs among key factors such as computational cost, latency, and generation quality pose significant challenges in semantic communications \cite{10999070}. Future 6G satellite communications must fulfill low-latency and highly reliable requirements, and the computation cost becomes an increasingly prominent challenge in intelligent future communication networks, particularly in satellite scenarios. In \cite{11207608}, a novel semantic efficiency metric was investigated to balance the tradeoffs between latency and generation quality in semantic communications for satellite networks, but the computational cost was not considered. Moreover, semantic efficiency metrics should be able to adapt to dynamic conditions and requirements under different tasks and environment limitations, which remains a challenging issue in existing studies.

Therefore, to fully study the potential of semantic communications in direct satellite-to-ground transmissions, and investigate the adaptive weight of factors in the semantic efficiency metric, we design an adaptive weighted evaluation metric under the semantic communication framework for direct satellite communications. The main contributions of this work are summarized as follows:
\begin{itemize}
  \item We introduce a semantic communication framework for direct satellite communications. To address challenges of stringent link budget, bursty service demands and low-latency constraints in satellite scenarios, we further introduce inter-satellite links (ISLs) to provide task migration among satellites, thereby improving transmission flexibility under link-budget limitations in satellite communications.

  \item We develop a decision-assisted deep reinforcement learning (DRL)-based optimization method to adapt the weights in the semantic efficiency metric based on task requirements and environment constraints. The proposed method enhances the sample efficiency in DRL, and balances computational cost, transmission latency, and semantic generation quality, to maximize the semantic communication efficiency in satellite-to-ground direct transmissions.

  \item Simulation results show the performance advantages of the semantic transmission framework in direct satellite communications, and verify the proposed adaptive weighting metric effectively balances robustness with efficiency.
\end{itemize}

The rest of this paper is organized as follows: Section \ref{se:sm} describes the system model, the proposed communication framework, the semantic efficiency metric, and the optimization problem. The proposed DRL-based method is introduced in Section \ref{sec:DRL}. Section \ref{sec:sim} presents the simulation results to verify the proposed framework and algorithm. Finally, Section \ref{sec:con} concludes this work.

\section{System Model and Problem Formulation}\label{se:sm}

\subsection{System Model}\label{subsec:sm_sys}
\begin{figure}[t!]
  \centering
  \centerline{\includegraphics[width=0.47\textwidth]{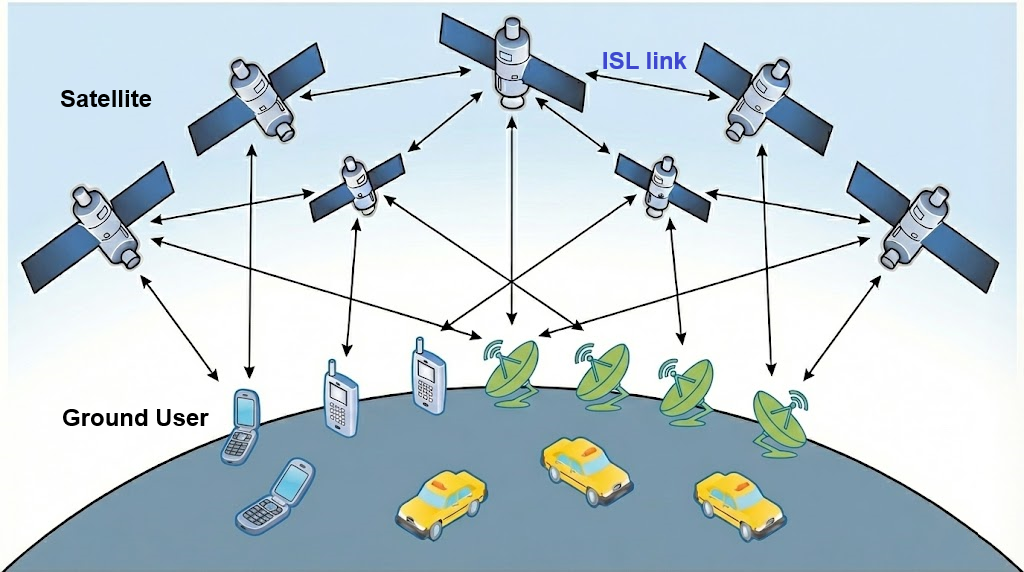}}
 \caption{Direct satellite communication networks.} \label{fig:SM}
\end{figure}
We consider a direct satellite communication system consisting of a low-Earth-orbit (LEO) satellite constellation and a set of ground users. The satellite set is denoted by $\mathcal{S}=\{S_1,S_2,\ldots,S_M\}$ and the ground user set is denoted by $\mathcal{U}=\{U_1,U_2,\ldots,U_N\}$. We consider the satellite coverage as in \cite{10440193}. We assume  $B_{m,n}$ is the bandwidth for the link between satellite $S_m$ and ground user $U_n$, where $m \in \{1,2,\ldots,M\}$ and $n \in \{1,2,\ldots,N\}$. Moreover, we introduce the satellite coverage with a minimum elevation angle $\aleph$ as in \cite{9344666}. At a given time instant, a satellite can serve only one user, and each user can be connected to only one satellite.

Following the communication model in \cite{9726800}, the channel coefficient from $S_m$ to $U_n$ at time slot $t$ is
\begin{equation}\label{eq:chan_hat}
\widehat{h}_{m,n}(t) = \frac{\sqrt{G_m}\,\lambda}{4\pi d_{m,n}(t)}e^{j\phi_{m,n}},
\end{equation}
where $G_m$ is the transmit antenna gain of satellite $S_m$, $\lambda$ is the carrier wavelength, and $\phi_{m,n}\in[0,2\pi)$ is the phase of satellite antenna. Considering channel aging due to Doppler and transmission delay, the outdated channel is expressed as
\begin{equation}\label{eq:chan_age}
h_{m,n}(t) = \rho_{m,n}\widehat{h}_{m,n}(t)+\sqrt{1-\rho_{m,n}^2}\,\widetilde{g}_{m,n}(t),
\end{equation}
where $\widetilde{g}_{m,n}(t)\sim\mathcal{CN}(0,1)$ has the same variance as the real CSI, and $\rho_{m,n}\in[0,1]$ measures the correlation between the outdated CSI and the real CSI.
$\rho_{m,n}(t) = J_0\!\left(2\pi f_{D,mn}\Delta\right)$, where $J_0(\cdot)$ is the $0$-th order Bessel function of the first kind, and $f_{D,mn}$ is the Doppler frequency. Thus, we can obtain the channel capacity for the link between $S_m$ and $U_n$ as
\begin{equation}\label{eq:rate_down}
R_{m,n}(t) = b_{m,n}\log_2\!\left(1+\frac{P_{S}\left|h_{m,n}(t)\right|^2}{ {\sigma}^2}\right),
\end{equation}
where $P_{S}$ denotes the transmit power at a satellite, $\sigma^2$ is the additive white Gaussian noise.

Moreover, satellites can forward tasks among themselves via ISLs when a satellite is out of the coverage. The ISL channel between satellites $S_{m1}$ and $S_{m2}$ is \cite{9327501}
\begin{equation}\small\label{eq:rate_ISL}
R_{m1 m2} = B_{m1 m2} {\rm{log_{2}}} \left( 1 +  \frac{P_{m1} |Y_{\rm max}|^2}{\imath \jmath B_{m1 m2} (\frac{{4 \pi d_{m1 m2} \ell_S}}{c})^2 } \right),
\end{equation}
where $B_{m1 m2}$ presents the bandwidth for the channel between $S_{m1}$ and $S_{m2}$, $P_{m1}$ denotes the ISL transmit power at satellite $S_{m1}$. $Y_{\rm max}$ is the peak gain of antennas of $S_{m1}$ in this direction, $d_{m1 m2}$ presents the distance between $S_{m1}$ and $S_{m2}$. $\imath$ denotes the Boltzmann constant, $\jmath$ is the thermal noise, $\ell_S$ presents the carrier frequency and $c$ denotes the speed of light.

Each user $U_n$ may generate a transmission task $\nu_{n(t)}$ at time slot $t$, and we assume the transmission can be completed within a time slot. We further extend the semantic efficiency metric formula in \cite{11207608} as
\begin{equation}\small\label{metric}
\begin{aligned}
\mu_{n(t)} = &~\kappa_{D, n(t)} (D_{\rm max, n(t)} - D_{n(t)}) + \kappa_{Q, n(t)} (Q_{n(t)} - Q_{\rm min, n(t)}) \\
&+ \kappa_{F, n(t)} ( F_n - F_{n(t)} ),
\end{aligned}
\end{equation}
where $D_{\rm max, n(t)}$ denotes the latency threshold for the task $\nu_n(t)$, and $Q_{\rm min, n(t)}$ denotes the semantic generation quality threshold, and $F_n$ denotes the computational resource of user $U_n$, $\kappa_{D, n(t)} \in [0.1, 0.9]$, $\kappa_{Q, n(t)}\in [0.1, 0.9]$ and $\kappa_{F, n(t)}\in [0.1, 0.9]$ are the weights in the metric for the task $\nu_{n(t)}$, $\kappa_{D, n(t)} + \kappa_{Q, n(t)} + \kappa_{F, n(t)} = 1$, $D_{n(t)}$ is the latency for transmission task $\nu_{n(t)}$, $Q_{n(t)}$ is the generation quality and $F_{n(t)}$ is the computational cost. It should be emphasized that three variables are normalized in \eqref{metric} by using the min–max scaling method. The weights in \eqref{metric} need to be optimized rather than fixed, so the metric is about intelligently balancing all three conflicting factors according to dynamic needs, thereby achieving an adaptive optimal balance among computational cost, latency, and generation quality. By jointly optimizing the semantic transmission modes and three weights of the semantic efficiency metric, the satellite communication system resources can be fully and effectively utilized to achieve true maximization.

\subsection{Semantic Transmission Framework}
We consider an image transmission framework which mixes conventional bit mode and semantic modes. Let $\mathbf{y}\in\mathbb{R}^{H\times W\times 3}$ denote the source data at the transmitter. The semantic encoder $\mathsf{E}_{\boldsymbol{\vartheta}}(\cdot)$ can map $\mathbf{y}$ into two features: a visual token sequence $\mathbf{c}$, and a textual vector $\mathbf{q}$ extracted from the same data. Thus, we can have three different transmission modes:

\begin{enumerate}
  \item Bit mode: Transmit all bits for the original data. This mode targets the highest pixel fidelity and lowest computational cost, but scales poorly with bandwidth.
  \item Text-only semantic mode: Transmit textual vector $\mathbf{q}$ only. The transmission cost is minimal, but the generation quality is not guaranteed.
  \item Hybrid semantic mode: Transmit $\mathbf{q}$ together with visual tokens $\mathbf{c}$. This mode requires more computational cost and transmission resource than Text-only semantic mode, but has higher generation quality.
\end{enumerate}
The above three modes present the choices in the proposed framework, but they can be further refined into sub-modes to provide high scalability. In this work, we define $\omega \in \{\omega_1, \omega_2, \omega_3, \omega_4, \omega_5\}$ as the transmission mode selection, where $\omega_1$ denotes bit mode, $\omega_2$ is the text-only semantic mode, others are sub-modes of hybrid semantic mode.

To support fine-grained scalability, the visual tokens are produced by a vector-quantized bottleneck.
Specifically, an image-to-latent mapper yields a continuous embedding map $\mathbf{e}\in\mathbb{R}^{A\times B\times d}$,
which is quantized to indices $\mathbf{c}\in\{1,\dots,C\}^{A\times B}$ using a codebook of size $C$.
The semantic payload size is $L_{\text{tok}} = AB\log_2 C \ \ \text{bits}$, and scalability can be implemented by adjusting $(A,B)$, pruning tokens, or layered packetization.

At the receiver, a semantic decoder $\mathsf{D}(\cdot)$ reconstructs an image by combining the global text cue $\mathbf{q}$ with the local token constraint $\mathbf{c}$.
We model $\mathsf{D}(\cdot)$ as a conditional generative process operating in a latent space.
Let $\mathbf{s}_0$ be the clean latent to be generated and $\mathbf{s}_\ell$ its progressively perturbed version at step $\ell\in\{1,\dots,L\}$: $\mathbf{s}_\ell = \sqrt{\overline{\chi}_\ell}\,\mathbf{s}_0 + \sqrt{1-\overline{\chi}_\ell}\,\boldsymbol{\nu}$, where $\boldsymbol{\nu}\sim\mathcal{N}(\mathbf{0},\mathbf{I})$ and $\overline{\chi}_\ell \triangleq \prod_{i=1}^{\ell}\chi_i$ is a predefined noise schedule.
Given the received conditions $(\mathbf{c},\mathbf{q})$, the latent is refined by a reverse update driven by a noise predictor $\widehat{\boldsymbol{\nu}}_{\boldsymbol{\vartheta}}(\cdot)$:
\begin{equation}\label{eq:rev_diff_new}
\mathbf{s}_{\ell-1}=\frac{1}{\sqrt{\chi_\ell}}
\left(\mathbf{s}_\ell-\frac{1-\chi_\ell}{\sqrt{1-\overline{\chi}_\ell}}\,
\widehat{\boldsymbol{\nu}}_{\boldsymbol{\vartheta}}(\mathbf{s}_\ell,\ell,\mathbf{c},\mathbf{q})\right).
\end{equation}
After iterating to $\mathbf{s}_0$, a lightweight latent-to-image mapper converts $\mathbf{s}_0$ into the final reconstruction $\widehat{\mathbf{y}}$.
This design naturally supports progressive quality: when only $\mathbf{q}$ is available the decoder produces a semantic-consistent sample, and when additional $\mathbf{c}$ packets arrive it sharpens structures and textures accordingly.

The training objective is to obtain (i) a discrete tokenization that is rate-efficient and (ii) a conditional generator that can exploit partial tokens.
We adopt a two-stage routine: tokenizer learning and conditional generator learning.

1. Vector-quantized tokenizer.
The encoder outputs $\mathbf{e}$ and is quantized to its nearest codeword $\mathbf{u}\in\mathbb{R}^{A\times B\times d}$ selected from a codebook $\{\mathbf{g}_1,\dots,\mathbf{g}_C\}$.
A standard commitment-style loss is used to align $\mathbf{e}$ and the selected codeword $\mathbf{u}$:
\begin{equation}\label{eq:vq_loss_new}
\mathcal{J}_{\text{VQ}}
=\mathbb{E}\!\left[\|\mathrm{sg}(\mathbf{e})-\mathbf{u}\|_2^2
+\upsilon\,\|\mathbf{e}-\mathrm{sg}(\mathbf{u})\|_2^2\right],
\end{equation}
where $\mathrm{sg}(\cdot)$ stops gradients and $\upsilon>0$ controls the commitment strength.
The token map $\mathbf{c}$ is then obtained by storing codeword indices, which directly determines the bit cost.

2. Conditional latent generator with token--text fusion.
We train the noise predictor $\widehat{\boldsymbol{\nu}}_{\boldsymbol{\vartheta}}(\cdot)$.
For each step $\ell$, we sample $\boldsymbol{\nu}$ and build $\mathbf{s}_\ell$ by the forward rule, then minimize the denoising error:
\begin{equation}\label{eq:denoise_loss_new}
\mathcal{J}_{\text{gen}}
=\mathbb{E}\!\left[\left\|\boldsymbol{\nu}-\widehat{\boldsymbol{\nu}}_{\boldsymbol{\vartheta}}(\mathbf{s}_\ell,\ell,\mathbf{c},\mathbf{q})\right\|_2^2\right].
\end{equation}
To make the decoder robust to packet losses and scalable reception, we randomly drop parts of $\mathbf{c}$ during training (e.g., masking token blocks or reducing $(A,B)$), forcing the model to interpolate missing local cues from $\mathbf{q}$ and the remaining tokens.
In deployment, one may further adjust semantic sharpness by blending conditional and unconditional predictions (using a learnable text embedding $\mathbf{q}_\varnothing$):
\begin{equation}\label{eq:guidance_new}
\widehat{\boldsymbol{\nu}} \leftarrow
(1+\zeta_g)\widehat{\boldsymbol{\nu}}_{\boldsymbol{\vartheta}}(\mathbf{s}_\ell,\ell,\mathbf{c},\mathbf{q})
-\zeta_g\,\widehat{\boldsymbol{\nu}}_{\boldsymbol{\vartheta}}(\mathbf{s}_\ell,\ell,\mathbf{c},\mathbf{q}_\varnothing),
\end{equation}
where $\zeta_g\ge 0$ is a guidance coefficient. In inference, the receiver runs a finite number of denoising iterations.
We denote by $L_{n(t)}\in\{1,2,\dots,10\}$ the selected number of denoising steps for task $\nu_{n(t)}$.
A larger $L_{n(t)}$ can provide higher generation quality but also bring higher computation cost.

Overall, the trained system can smoothly trade payload for fidelity: when bandwidth is limited, transmit only $\mathbf{q}$ or a small token subset; when resources permit, send more tokens to progressively approach near-lossless perceptual quality.

\subsection{Problem Formulation}
In direct satellite networks, we aim to enhance the average semantic efficiency metric by jointly optimizing the semantic transmission mode selection, ISL selection , denoising steps, semantic efficiency metric weights $\kappa_{n(t)} = \{\kappa_{D, n(t)}, \kappa_{Q, n(t)}, \kappa_{F, n(t)}\}$, and satellite-user association $\chi_{m,n}(t)$, where $\chi_{m,n}(t) = 1$ indicates that $S_m$ can transmit data to $U_n$ at time slot $t$, otherwise $\chi_{m,n}(t) = 0$. The problem formulation is as

\begin{align}
    &\max_{\omega_{n(t)}, m1_{n(t)}(t), m2_{n(t)}(t), L_{n(t)}, \kappa_{n(t)}, \chi_{n(t)}} \frac{1}{TN}\sum_{t=1}^{T}\sum_{n=1}^{N}\mu_{n(t)},\label{SecrecyFunc}\\
    ~~~~~~&{\rm s.t.}~ \sum_{n=1}^{N} \chi_{m,n}(t) \in \{0, 1\}, \forall m \in {\mathcal M} \tag{\ref{SecrecyFunc}{a}}, \label{SecrecyFuncSuba}\\
    &~~~~~~\sum_{m=1}^{M} \chi_{m,n}(t) \in \{0, 1\}, \forall n \in {\mathcal N} \tag{\ref{SecrecyFunc}{b}}, \label{SecrecyFuncSubb}\\
    &~~~~~~m1_{n(t)}(t) \neq m2_{n(t)}(t)  \tag{\ref{SecrecyFunc}{c}}, \label{SecrecyFuncSubc}\\
    &~~~~~~D_{\rm max, n(t)} \geq D_{n(t)} \tag{\ref{SecrecyFunc}{d}}, \label{SecrecyFuncSubd}\\
    &~~~~~~Q_{\rm min, n(t)} \leq Q_{n(t)} \tag{\ref{SecrecyFunc}{e}}, \label{SecrecyFuncSube}\\
    &~~~~~~F_n \geq F_{n(t)} \tag{\ref{SecrecyFunc}{f}}, \label{SecrecyFuncSubf}
\end{align}
where $T$ is the number of total time slots. \eqref{SecrecyFuncSuba} and \eqref{SecrecyFuncSubb} denote one satellite can only serve one ground user, and one ground user can access only one satellite. \eqref{SecrecyFuncSubc} is the ISL constraint. \eqref{SecrecyFuncSubd} denotes the latency requirement of task $\nu_{n(t)}$, \eqref{SecrecyFuncSube} presents the generation quality requirement, and \eqref{SecrecyFuncSubf} is the computation resource limitation for user $U_n$. In \eqref{SecrecyFunc}, it is a complex nonlinear integer programming problem that requires long-term optimization over $T$ time slots. Moreover, considering the task heterogeneity and burstiness, as well as the dynamic satellite networks, the optimization must be performed in a dynamic environment rather than a static system. Therefore, we introduce a DRL algorithm to address this problem.

\section{Decision-Assisted DRL-Based Optimization}\label{sec:DRL}
The optimization problem in \eqref{SecrecyFunc} is a complex nonlinear integer programming problem. To achieve fast decision making with dynamic task arrivals, we introduce the REINFORCE++ \cite{hureinforceplusplus} which is a novel DRL method.
In a Markov decision process, we define the state as $s(t) = \{(h_{m,n}(t), \forall m \& \forall n), (D_{\rm max, n(t)}, Q_{\rm min, n(t)}, F_n, \forall n) \}$  at time slot $t$. The action is defined as $\mathbf{a}(t) = \{\omega_{n(t)}, m1_{n(t)}(t), m2_{n(t)}(t), L_{n(t)}, \kappa_{n(t)}, \chi_{n(t)}\}$. We define the instantaneous reward as the average semantic efficiency $r(t)=\frac{1}{N_1(t)}\sum_{n=1}^{N_1}\mu_{n(t)}$, where $N_1(t)$ denotes the number of arrived tasks at time slot $t$, and the objective is to maximize the expected discounted return $\mathbb{E}\!\left[\sum_{t=1}^{T}\gamma^{t-1}r(t)\right]$, where $\gamma$ denotes the discount factor. In REINFORCE++, the method normalizes advantages using global batch statistics:
\begin{equation}\label{eq:global_norm_rfpp}
\widehat{A}_t=\frac{G_t-\operatorname{mean}_{\mathcal{B}}(G)}{\operatorname{std}_{\mathcal{B}}(G)+\epsilon},
\end{equation}
where $\mathcal{B}$ denotes the entire sampled batch, and $\epsilon$ is a small constant.
Then, similar to a PPO design without critic networks, REINFORCE++ maximizes the clipped surrogate objective as
\begin{equation}\label{eq:clip_obj_rfpp}
\mathcal{L}_{\rm clip}(\theta)=
\mathbb{E}_{t\sim\mathcal{B}}\!\left[
\min\!\big(\rho_t\widehat{A}_t,\ \operatorname{clip}(\rho_t,1\!-\!\varepsilon,1\!+\!\varepsilon)\widehat{A}_t\big)
\right],
\end{equation}
with $\rho_t=\pi_\theta(a_t|s_t)/\pi_{\theta_{\rm old}}(a_t|s_t)$.
To prevent policy drift, REINFORCE++ employs an explicit KL penalty implemented via the $k2$ estimator (a KL Approximation) as
\begin{equation}\label{eq:kl_k2_rfpp}
\mathcal{J}_{k2}(\theta)=\mathbb{E}_{t\sim\mathcal{B}}\!\left[\frac{1}{2}\Big(\log\frac{\pi_\theta(a_t|s_t)}{\pi_{\rm ref}(a_t|s_t)}\Big)^2\right],
\end{equation}
where $\pi_{\rm ref}$ is a fixed reference policy (e.g., the initial policy).
Finally, we update $\theta$ by ascending $\mathcal{L}_{\rm clip}(\theta)-\lambda \mathcal{J}_{k2}(\theta)$. Due to space limitations, the details can be found in \cite{hureinforceplusplus}.

To further improve the sample efficiency in DRL, we introduce a decision assistant that utilizes a-priori feasibility information to reduce invalid exploration \cite{10440193}. The main idea is to identify an action-feasible set $\mathcal{A}_{\rm feas}(s_t)\subseteq\mathcal{A}$ from deterministic constraints in \eqref{SecrecyFunc} (e.g., out of the satellite coverage), and then restrict the policy sampling to this set.

To be specific, we apply action masking on the policy logits: for any $a\notin\mathcal{A}_{\rm feas}(s_t)$, its logit is set to $-\infty$ before the softmax so that $\pi_\theta(a|s_t)=0$, and the remaining probabilities are renormalized. This method narrows the range of the original large action space into a smaller feasible one, therefore, this method can significantly reduce the variance in policy-gradient updates. Notice that due to the space limitation, other technical details can refer to the original algorithm paper \cite{hureinforceplusplus}.

\begin{algorithm}[t]
\caption{Decision-Assisted REINFORCE++ Optimization}
\label{alg1}
\small
\begin{algorithmic}[1]
\State Initialize policy $\pi_\theta$ and set reference policy $\pi_{\rm ref}$.
\For{each iteration}
\State Set $\pi_{\theta_{\rm old}}\leftarrow \pi_\theta$; collect a batch of trajectories and invalid
\Statex \phantom{0}\phantom{0}\phantom{0}samples using $\pi_{\theta_{\rm old}}$.
\State Compute rewards $r_t$ and returns $G_t$.
\State Compute globally normalized advantages $\widehat{A}_t$ by \eqref{eq:global_norm_rfpp}.
\State Compute $\rho_t=\pi_\theta(a_t|s_t)/\pi_{\theta_{\rm old}}(a_t|s_t)$.
\State Update $\theta$ by gradient ascent on $\mathcal{L}_{\rm clip}(\theta)-\lambda \mathcal{J}_{k2}(\theta)$ using
\Statex \phantom{0}\phantom{0}\phantom{0}\eqref{eq:clip_obj_rfpp} and \eqref{eq:kl_k2_rfpp}.
\EndFor
\end{algorithmic}
\end{algorithm}

\section{NUMERICAL RESULTS}\label{sec:sim}
The key simulation parameters are summarized: the number of satellites $M = 8$, the number of ground users $N = 5$, ground users are randomly distributed on the same side of the Earth, the altitude of satellites is 700 km, the velocity of satellites is 7.5 km/s, the gain at satellite antennas is 40 dB, the Doppler frequency is the same as in Ka-band, the carrier frequency is 25 GHz, the transmit power of satellites is 1 W, the bandwidth is 30 MHz, tasks arrive as in 3GPP FTP Model 3 for each user, the size of task is 3.5 MB, the latency thresholds are generated between 0.5 s and 2 s, the generation quality thresholds are between 13 and 21 (PSNR), the computational resource of a ground user is between 700 GFlops and 7000 GFlops, the computational cost for a denoising step is 686 GFlops, other communication and semantic transmission settings can be found in \cite{11207608}.

\begin{figure}[t!]
  \centering
  \centerline{\includegraphics[scale=0.56]{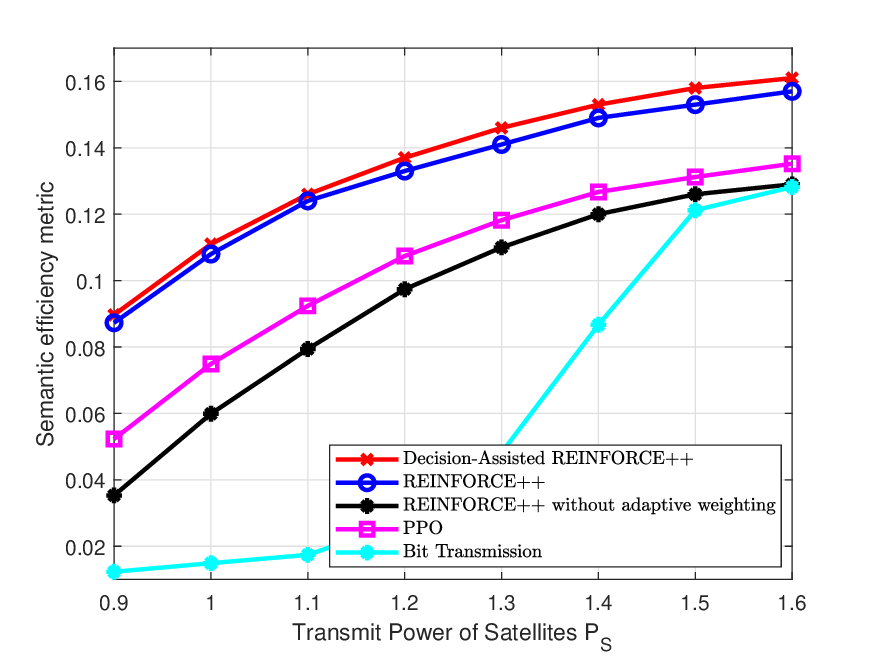}}
 \caption{\small Average semantic efficiency versus transmit power at satellites.} \label{fig:R1}
\end{figure}
Fig. \ref{fig:R1} shows the variation of semantic communication efficiency as the satellite transmit power increases. The proposed decision-assisted REINFORCE++ algorithm clearly outperforms PPO, REINFORCE++ without adaptive weighting, and REINFORCE++ without decision-assistant, and shows weight optimization having a substantial impact on semantic communication efficiency. Moreover, when the satellite transmit power is high, the link budget is sufficient, the impact of transmission delay in bit communications is significantly reduced to enhance the efficiency metric. However, direct satellite communications typically suffer from stringent link budget constraints, under which traditional bit transmissions become inefficient due to huge latency. In contrast, the proposed semantic framework exhibits clear advantages in direct satellite communication scenarios.

\begin{figure}[t!]
  \centering
  \centerline{\includegraphics[scale=0.56]{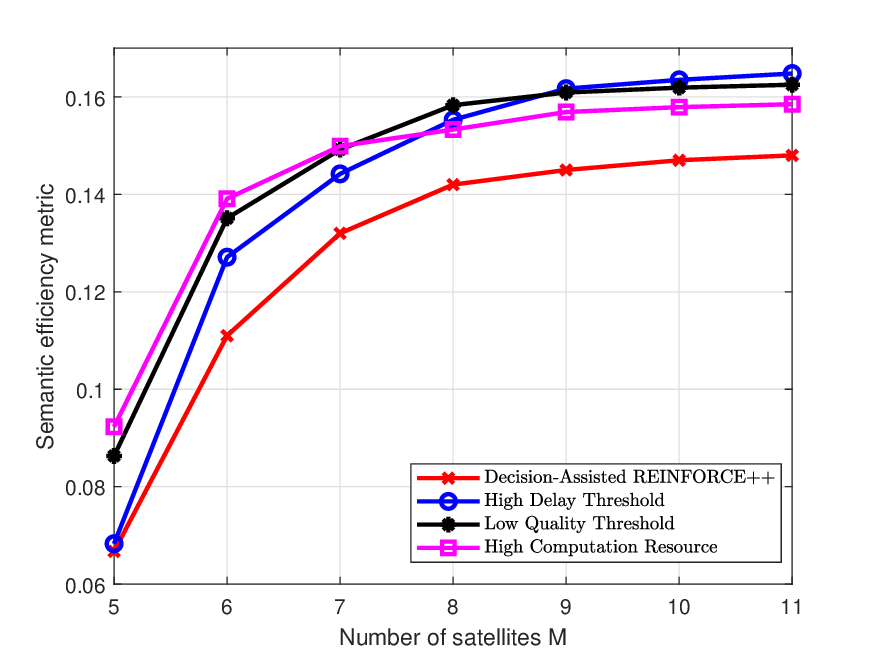}}
 \caption{\small Average semantic efficiency versus number of satellites.} \label{fig:R2}
\end{figure}
Fig. \ref{fig:R2} illustrates the impact of the number of satellites on direct satellite communication performance. When the number of satellites is small, transmissions and ISL forwarding cannot always be completed in time due to limited satellite coverage. To satisfy latency requirements, it may be necessary to sacrifice the generation quality and computational resources, resulting in a lower overall performance metric. As the number of satellites increases, satellites can provide nearly seamless coverage, so the semantic communication efficiency increases rapidly and then gradually stabilizes. When we vary the latency thresholds, quality thresholds, and ground user computational constraints of tasks, respectively, it can be observed that with a larger number of satellites, latency constraints become relatively relaxed. Thus, tasks with higher latency threshold can achieve greater performance gains. Conversely, when the number of satellites is small, the link budget is limited, achieving higher performance relies more heavily on computational resources to enhance generation quality. In this case, users with higher computational capability can obtain higher efficiency values.

\section{Conclusion}\label{sec:con}
This paper proposed a semantic communication framework to address the link budget challenge for direct satellite communications. By introducing three hybrid transmission modes and three sub-modes, a denoising-step control mechanism, and an adaptive weighting semantic efficiency metric, the proposed framework offers tradeoffs in transmission size, computation cost, and generation quality. The optimization problem was addressed via a decision-assisted REINFORCE++ DRL algorithm, which improves sample efficiency by introducing infeasible samples and stabilizes critic-free policy learning. Simulation results verified the effectiveness of the decision-assisted framework and the adaptive weighting design in the proposed algorithm, and the impact of the number of satellites, as well as the impact of transmit power. This work presents a viable and effective solution that fundamentally enables reliable direct satellite-to-user access under stringent link budgets, a scenario where conventional bit-level satellite communication would typically fail. It marks a significant step toward making ubiquitous, user-centric satellite connectivity a practical reality.


\bibliographystyle{ieeetr}
\bibliography{ref}


\end{document}